\documentclass[a4paper,11pt]{article} % option doublespacing or reviewcopy for double line spacing

\usepackage{graphicx} % for more complicated commands
\usepackage{amsmath,amsfonts,amssymb,bm}
\usepackage{color}
\usepackage{epstopdf}
\usepackage{subfig} 
\usepackage[utf8]{inputenc}

\usepackage{microtype} % for ligatures  ff fl fi, etc
\DisableLigatures{encoding = *, family = *}

\usepackage{ragged2e} % has command \justifying

\usepackage{fancyhdr}
\fancypagestyle{plain}{\pagestyle{fancy}} % adds on the first page as well

\usepackage{blindtext}

\usepackage[sort&compress,numbers]{natbib}
\usepackage{doi}%<----------

\begin{document}

\title{Magnetic resonance measurement and reconstruction of the diffusion propagator}

%\justifying

\author{Alfredo Miguel Ordinola Santisteban \& Evren \"Ozarslan$^\dagger$\\ %, Second Authos$^b$  \\
	{}\footnotesize\it Department of Biomedical Engineering, Campus US, Linköping University, SE-581 85 Linköping, Sweden. \\
\footnotesize\it  ${}^\dagger$E-mail: evren.ozarslan@liu.se} %\\
\date{}

\maketitle
\pagestyle{fancy}

\begin{abstract}
We demonstrate the experimental determination of the diffusion propagator using magnetic resonance (MR) techniques. To this end, a recently introduced method was implemented on a benchtop MR scanner and incorporated into profiling and imaging sequences. Diffusion encoding was obtained by three gradient pulse pairs, each applied around a respective 180$^\circ$ radiofrequency pulse. The data involving two independent wavenumbers were transformed from the measurement domain to the spatial domain, yielding an apparent diffusion propagator. For free diffusion, this apparent propagator converges to the true one when two of the pulse pairs are replaced with pulses of infinitesimal duration. 
\end{abstract}

%%%%%%%%%%%%%%%%%%%%%%%%%%%%%%%%%%%%%%%%%%%%%%%%%%%%%%%%%%%%%%%
\section{Introduction}

\begin{figure}[t!]
	%\centering
	\subfloat[]{
		\includegraphics[width=0.6\textwidth]{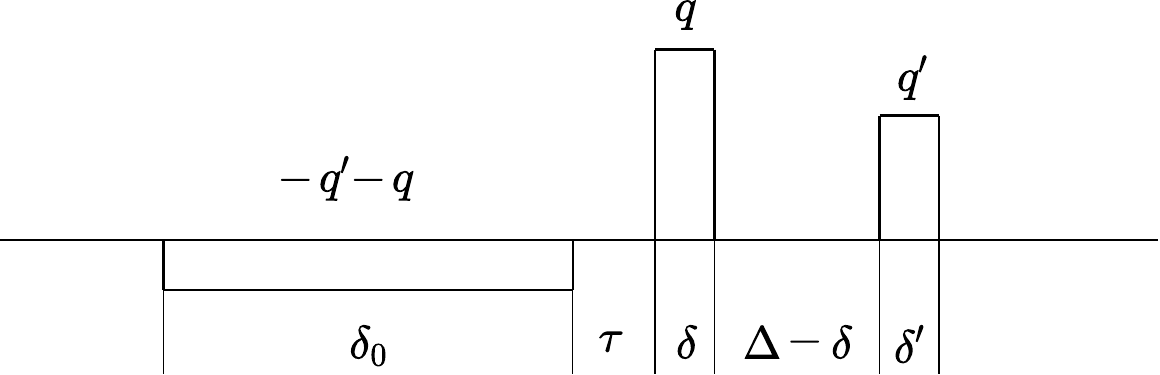}\label{fig:P_sequence}} % label used to refer to the first image, will appear as 1a
	\\[30pt]%\vspace{14pt}
	\subfloat[]{
		\includegraphics[width=0.98\textwidth]{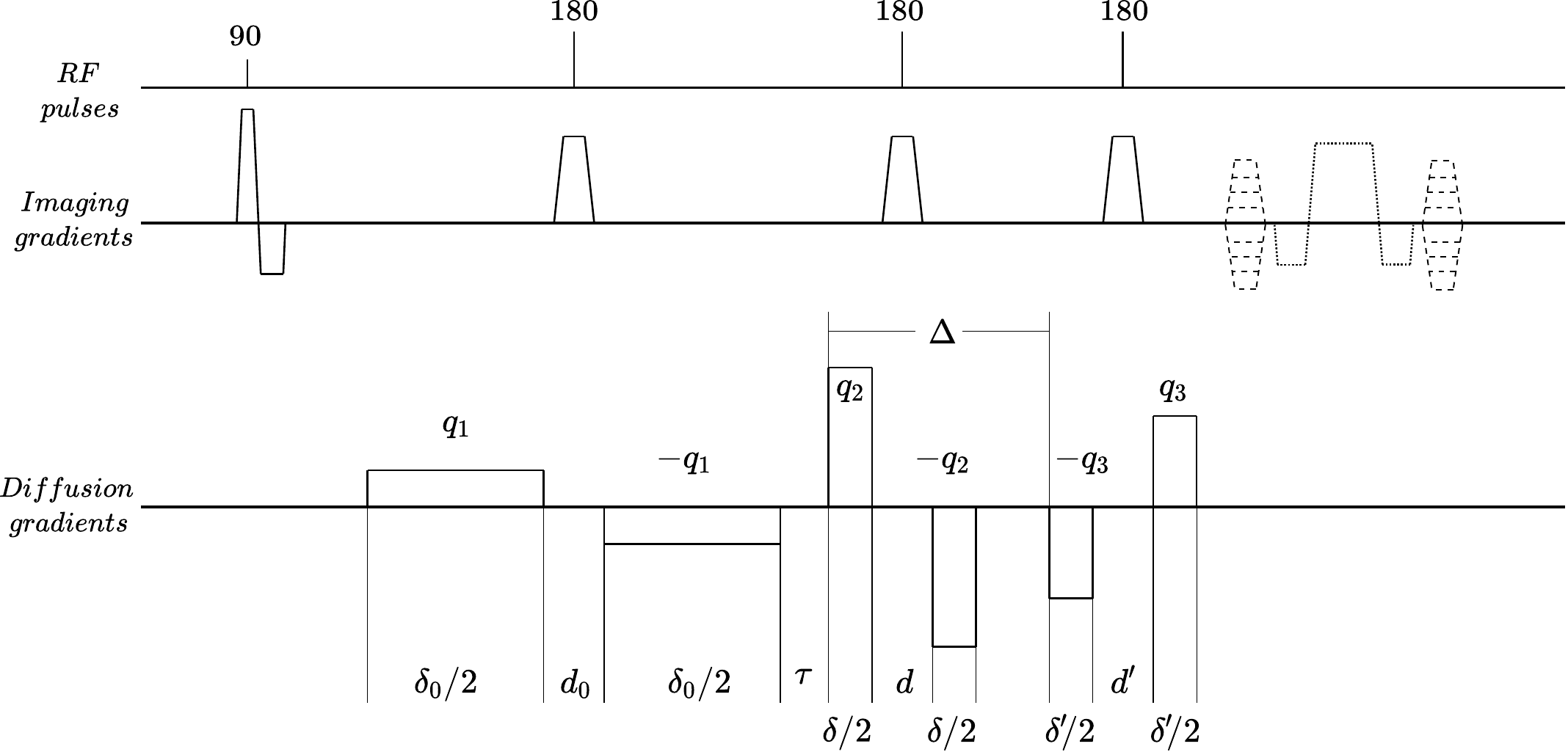}\label{fig:BP_sequence}} % label used to refer to the second image, will appear as 1b
	\vspace{14pt}\caption{(a) Effective gradient waveform introduced in \cite{Ozarslan21ISMRMeverything,Ozarslan21ARXIVeverything}. (b) Implemented sequence with bipolar gradients, where the areas under the diffusion gradient pulses are: $q_1 = \frac{q+q'}{2}$, $q_2 = \frac{q}{2}$ and $q_3 = \frac{q'}{2}$, respectively. In our implementation, the delays between the two lobes of the bipolar gradients ($d_0$, $d$, and $d'$) are the same for all three pulse pairs. Imaging gradients in the second row enable slice selection (straight line), frequency encoding (dotted line), and phase encoding (dashed line). \label{fig:sequence}}
	% label used to reference the entire figure
\end{figure}

Recently, one of us introduced a novel MR diffusion encoding method that could be employed to determine the diffusion propagator \cite{Ozarslan21ISMRMeverything,Ozarslan21ARXIVeverything}. In its most concise form, the technique features three gradient pulses two of which are determined independently. Such a sequence is illustrated in Figure \ref{fig:P_sequence} and it should be noted that it reduces to the classic Stejskal-Tanner sequence \cite{StejskalTanner65} for $\mathbf q'=-\mathbf q$ and to Laun et al.'s sequence \cite{Laun11prl} for $\mathbf q' = \mathbf 0$.

To illustrate how this waveform functions, let us consider all of the Brownian trajectories within the specimen and first focus on the collection of trajectories having the same mean position during the application of the first pulse. For simplicity, we refer to such a collection as a ``bouquet.'' Thus, by ``a bouquet at $\mathbf x_1$,'' we refer to the collection of paths whose mean position during the application of the first pulse is $\mathbf x_1$, and the distribution of bouquets is denoted by $\beta(\mathbf x)$. The effect of the first pulse is then to modulate the magnetization for the bouquet by a factor given by $e^{i (\mathbf q + \mathbf q')\cdot \mathbf x_1}$, where $\mathbf q$ and $\mathbf q'$ are the temporal integrals of, respectively, the second and third gradient vectors multiplied by the gyromagnetic ratio. 

Let $\mathbf x_2$ and $\mathbf x_3$ denote the mean positions of the particles during the second and third pulses, respectively, leading to modulation factors of $e^{-i \mathbf q \cdot \mathbf x_2}$ and $e^{-i \mathbf q' \cdot \mathbf x_3}$. The overall $\mathbf q$-dependent effect is then $e^{-i \mathbf q \cdot (\mathbf x_2-\mathbf x_1)}$ while the same for $\mathbf q'$ is $e^{-i \mathbf q' \cdot (\mathbf x_3-\mathbf x_1)}$. With the definitions $\mathbf x = \mathbf x_2 -\mathbf x_1$ and $\mathbf x' = \mathbf x_3 -\mathbf x_1$, the signal attenuation is %for a bouquet at $\mathbf x_1$ is 
$$E_\Delta(\mathbf{q},\mathbf{q}')=\int \mathrm{d} \mathbf{x_1} \, \beta(\mathbf{x_1}) \int \mathrm{d} \mathbf{x} \, \rho(\mathbf x_1+ \mathbf{x} | \mathbf x_1) \int \mathrm{d}\mathbf{x}' \, P(\mathbf x_1+\mathbf{x}', \Delta | \mathbf x_1+\mathbf{x}) \, e^{-i (\mathbf{q}\cdot\mathbf{x}+\mathbf{q}'\cdot\mathbf{x}' )}\ ,$$
where $\rho(\mathbf x_1+ \mathbf{x} | \mathbf x_1)$ denotes the probability density for finding a particle of bouquet at $\mathbf x_1$ to have the mean position $\mathbf x_1+ \mathbf{x}$ during the application of the second pulse. 

In this work, we demonstrate the application of the method for the case of free diffusion. In this case, $\beta(\mathbf{x_1})$ is a constant while $\rho(\mathbf x_1+ \mathbf{x} | \mathbf x_1)$ and $P(\mathbf x_1+\mathbf{x}', \Delta | \mathbf x_1+\mathbf{x})$ are translationally-invariant. Thus, the signal attenuation is
$$E_\Delta(\mathbf{q},\mathbf{q}')=\int \mathrm{d} \mathbf{x} \, \rho(\mathbf{x} | \mathbf 0) \int \mathrm{d}\mathbf{x}' \, P(\mathbf{x}', \Delta | \mathbf{x}) \, e^{-i (\mathbf{q}\cdot\mathbf{x}+\mathbf{q}'\cdot\mathbf{x}' )}\ ,$$
and the diffusion propagator is given by \cite{Ozarslan21ISMRMeverything,Ozarslan21ARXIVeverything}
\begin{align} \label{eq:P_from_E}
P(\mathbf{x}',\Delta|\mathbf{x}) & =\frac{\int\mathrm{d}\mathbf{q}\, e^{i\mathbf{q}\cdot\mathbf{x}}\int\mathrm{d}\mathbf{q}'\, e^{i\mathbf{q}'\cdot\mathbf{x}'}\, E_\Delta(\mathbf{q},\mathbf{q}')}{(2\pi)^n \int\mathrm{d}\mathbf{q}\,e^{i\mathbf{q}\cdot\mathbf{x}}\, E_\Delta(\mathbf{q},\mathbf{0})}\ ,
\end{align}
where $n$ is the number of dimensions of the space in which the measurement is performed.

Here, we present experimental measurements and reconstructions of the diffusion propagator. The implemented sequence is based on the previous work presented in \cite{Ozarslan21ISMRMeverything,Ozarslan21ARXIVeverything}. However, in our implementation, each diffusion gradient is replaced with bipolar gradients as shown in Figure \ref{fig:BP_sequence}. This diffusion encoding scheme is incorporated into two sequences. In the first, only selective excitation and frequency encoding are utilized, thus yielding a projection of the sample; we refer to such acquisitions as profiling measurements. The second sequence employs phase encoding in order to acquire diffusion propagator valued images of the specimen.

%%%%%%%%%%%%%%%%%%%%%%%%%%%%%%%%%%%%%%%%%%%%%%%%%%%%%%%%%%%%%%%
\section{Theory}

To validate the results obtained from the proposed acquisition sequence, a simulation with the same parameters as the imaging experiment can be performed. To this end, the simulated or ``ideal'' signal decay from the proposed pulse sequence can be used to reconstruct the propagator with the same method used for the experimental data. In our experiments, all diffusion encoding gradients are applied along the same orientation, i.e., $n=1$. All experimental parameters are defined in Figure \ref{fig:BP_sequence}. For simplicity, we ignore the finite slew rate, and represent all gradient pulses with rectangles having the same area as the underlying trapezoidal pulses.

The analytical expression for the signal attenuation is calculated using an expression derived by Karlicek and Lowe \cite{Karlicek80}, and is obtained to be
\begin{align}
E(q,q') &= \exp \left\{ -\frac{D_0}{12} \left[(2 \mu+4\delta+3d) q^2 + 4 (\mu+3\delta+3d) qq' + (2 \mu+12\Delta+4\delta'+3d') q'^2 \right] \right\} \ ,
\end{align}
where $\mu=2\delta_0+6\tau+3d_0/2$. Employing this expression in equation \ref{eq:P_from_E}, one obtains the  apparent propagator
\begin{align}
P_\mathrm{app}(x',\Delta | x) &=  (4\pi D_0 T )^{-1/2} \, e^{-(x' - a x)^2 / (4 D_0 T)} \label{eq:P_app} \ ,
\end{align}
where
\begin{align}
a &= \frac{2 \mu + 6\delta+6d}{2 \mu + 4\delta+3d} \ , \textrm{ and} \nonumber\\[10pt]
T &= \frac{1}{12}\left( 12\Delta+4\delta'+3d' - \frac{2\mu (8\delta + 9 d) + 36 (\delta + d)^2}{2\mu + 4\delta + 3 d}   \right) \ .
\end{align}

As $d \rightarrow 0$ , $\delta \rightarrow 0$, $d' \rightarrow 0$ , and $\delta' \rightarrow 0$, the apparent propagator approaches the true propagator $P(x',\Delta | x) = (4\pi D_0 \Delta )^{-1/2} \, e^{-(x' - x)^2 / (4 D_0 \Delta )} $ as expected. The deviation from this form is due to the finite value of these durations. On the other hand, having longer $\delta_0$, $d_0$, and $\tau$ is advantageous in order for the apparent propagator to provide a good approximation to the true propagator. One consequence of not having $d=\delta=0$ is the deviation of $a$ from unity. This results in an apparent violation of the reciprocity condition $P(x',\Delta | x) = P(x,\Delta | x')$. Thus, one should not mistake this for the presence of an external force \cite{Ozarslan21ISMRMeverything,Ozarslan21ARXIVeverything}. Another consequence of having nonzero $d$ or $\delta$ is that the integral of $P_\mathrm{app}(x',\Delta | x) $ over $x$ is $a^{-1}$, thus it deviates from its expected value of 1. Both of these issues result from deviations from the `ideal' experimental conditions, and could be remedied via redefining $x$ as $ax$. Figure \ref{fig:Panalytical} illustrates the true and apparent propagators as well as how such rescaling of $x$ addresses their discrepancy.

%%%%%%%%%%%%%%%%%%%%%%%%%%%%%%%%%%%%%%%%%%%%%%%%%%%%%%%%%%%%%%%
\section{Methods}

The sequence and experiments presented in this study were implemented and performed on a benchtop MRI scanner (\emph{Pure Devices GmbH, Germany}). It should be noted that the employed scanner's bore is placed along the $y$-axis of its reference frame. Thus, in order to obtain profiles and images of the sample's cross-section perpendicular to the bore, imaging gradients were applied as follows: slice selection along the $y$-axis, phase encoding along the $x$-axis, and frequency encoding along the $z$-axis.

For this particular application, the separation between bipolar gradients was kept constant along all three pairs, which are symmetrically placed around the {180$^\circ$} pulses. Moreover, the diffusion time $\Delta$ is defined to be the delay between the beginning of the first pulse of the second pair and the beginning of the first pulse of the third pair.
 
As the propagator has infinite support, the reconstruction through the previously presented inverse Fourier transform is expected to exhibit aliasing effects. To alleviate this issue, the signal decay was ``oversampled'', i.e., more values of $q$ and $q'$ were sampled in order to obtain a larger $x$ and $x'$ space after reconstruction. Only the center of the reconstructed propagator, with a number of samples depending on the oversampling factor, is then obtained to suppress the aforementioned aliasing effects. 

Two experiments were performed on a water phantom for this proof-of-concept study. From these, three datasets were acquired and processed to reconstruct the diffusion propagator. In the first experiment, phase encoding gradients were turned off in order to effectively obtain the profile of the sample. This can be described as the projection of the measured slice onto the frequency encoding direction. Pulse durations $\delta_0$, $\delta$ and $\delta'$ were set to \mbox{20 ms}, \mbox{4 ms} and \mbox{4 ms}, respectively, with a diffusion time $\Delta$ of 10 ms. The separation time between the first and second gradient pulse pairs $\tau$ was set to 3 ms and the separation times between bipolar gradient pairs $d_0$, $d$ and $d'$ were all set to \mbox{4 ms}. The values of $q$ and $q'$ were varied from $-1$ rad/$\mu$m to 1 rad/$\mu$m with 31 linearly spaced values, yielding a total of 961 measurements. Furthermore, the center value in k-space of these measurements was used in order to obtain the propagator in a spectroscopy acquisition. The effective echo time (TE) of the sequence resulted in 58 ms and the repetition time (TR) was set to 2 s, yielding a total acquisition time of approximately 1.7 hours.

For the second experiment, images of the water phantom were obtained with a Field-of-View (FOV) of \mbox{$10 \times 10$ mm$^2$} and a \mbox{$16 \times 16$} matrix. The time parameters $\delta_0$, $\delta$, $\delta'$ $\Delta$, $\tau$, $d_0$, $d$ and $d'$ were set to the same values as in the previous experiment. The values of $q$ and $q'$ were varied from $-1.2$ rad/$\mu$m to 1.2 rad/$\mu$m with 21 linearly spaced values, yielding a total of 441 measurements. The TE and TR were 58 ms and \mbox{2 s}, respectively, yielding a total acquisition time of approximately 8.6 hours. The acquisition was repeated three times in the same phantom and averaged prior to data processing. The oversampling factor was set to 2 and all diffusion gradients were applied along the $x$ axis of the scanner's reference frame in both experiments.

The acquired data were normalized with the unweighted signal in each experiment, i.e., $S_\Delta(0,0)$, to obtain the signal attenuation $E_\Delta(q,q')$, and the propagator $P(x',\Delta|x)$ was reconstructed following equation \ref{eq:P_from_E}. It should be noted that the reconstruction was obtained through the quotient of the double integral in the numerator and single integral in the denominator. This was done with the goal of computing the full inverse Fourier transform and avoiding the use of a built-in inverse fast Fourier transform function in either case. All computations were performed using MATLAB (\emph{Mathworks Inc., Natick, MA, USA}).

To asses the quality of the experimentally reconstructed propagator, difference maps between the latter and both the analytical expression for the apparent propagator and the one obtained through simulated data were calculated.

Furthermore, it is expected to observe low SNR and Rician distributed signal in the acquired data, which will prevent the signal to fully decay at high values of $q$ and $q'$. For this reason, noisy signal decay profiles were synthesized  through the expression
$$\hat{S}_\Delta(q,q')=\sqrt{(S_\Delta(q,q') + n_1(0,\sigma))^2+n_2(0,\sigma)^2} ,$$
where $n_1$ and $n_2$ are two independent Gaussian noise sources with mean zero and standard deviation $\sigma$. The propagator could then be reconstructed from $\hat{E}_\Delta(q,q')$ and compared to the obtained results in this study. The $\sigma$ parameter was arbitrarily set to 0.01 in the presented simulation in order to qualitatively analyze the effect of this type of noise in the proposed framework.

%%%%%%%%%%%%%%%%%%%%%%%%%%%%%%%%%%%%%%%%%%%%%%%%%%%%%%%%%%%%%%%
\section{Results}

\begin{figure}[h!]
	\begin{center}
		\includegraphics[width=\textwidth]{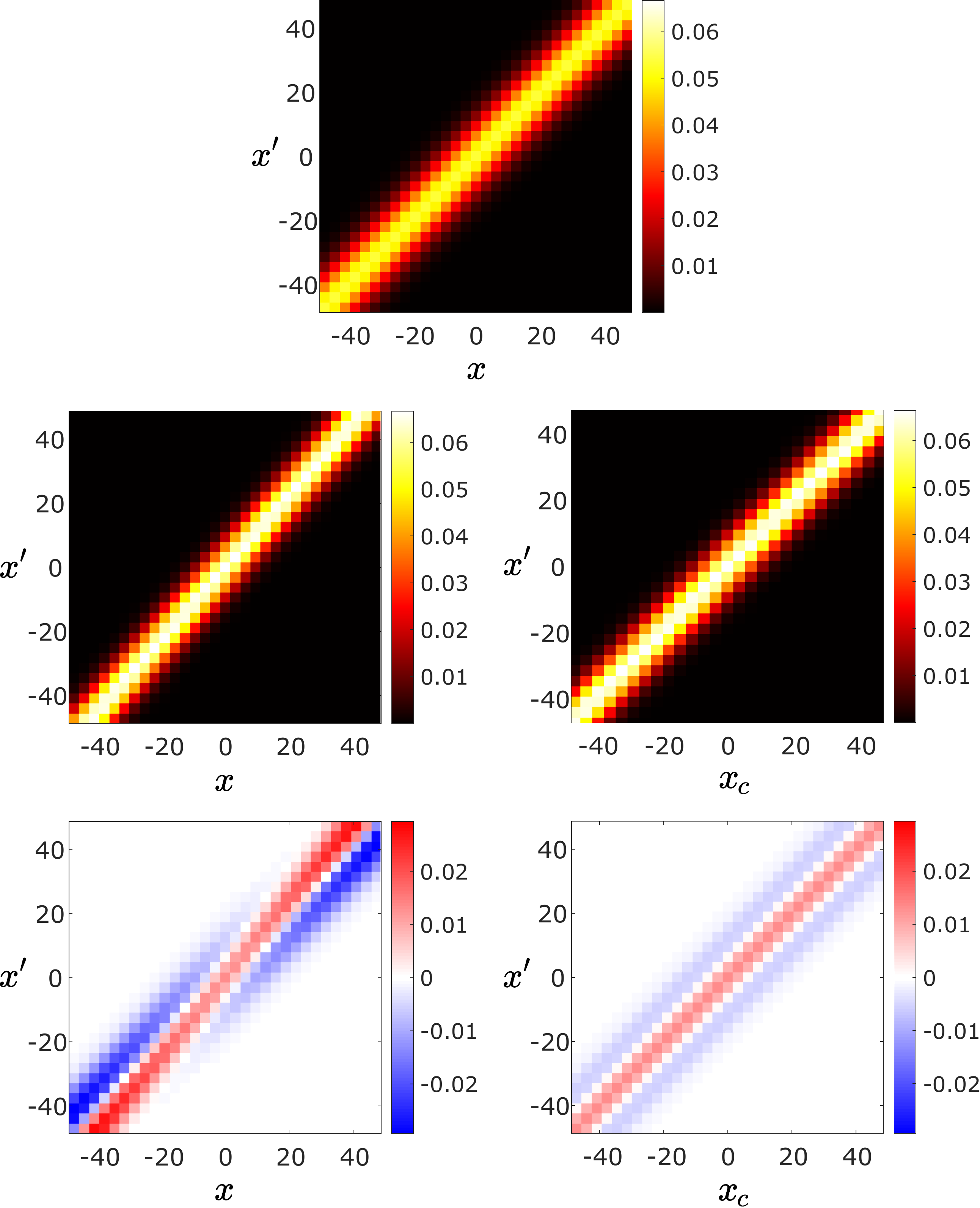}
	\end{center}
	\vspace{-14pt}\caption{Upper row: True propagator i.e., when $d \rightarrow 0$, $\delta \rightarrow 0$, $d' \rightarrow 0$ and $\delta' \rightarrow 0$. Middle row: Apparent propagator (left) and  apparent propagator where the $x$ axis has been corrected according to $x_c = ax$. Bottom row: Difference maps between the apparent propagator and true propagator:  $P_\mathrm{app}(x',\Delta | x)-P_\mathrm{true}(x',\Delta | x)$ (left) and $P_\mathrm{app}(x',\Delta | x_c)-P_\mathrm{true}(x',\Delta | x)$ (right). \label{fig:Panalytical}}
\end{figure}

The true propagator, the apparent propagator and the apparent propagator corrected with the aforementioned change in $x$ axis, are presented in Figure \ref{fig:Panalytical}. Furthermore, the difference maps between both apparent propagators and the true one are presented in the bottom row of said figure.

\begin{figure}[h!]
	\begin{center}
		\includegraphics[width=\textwidth]{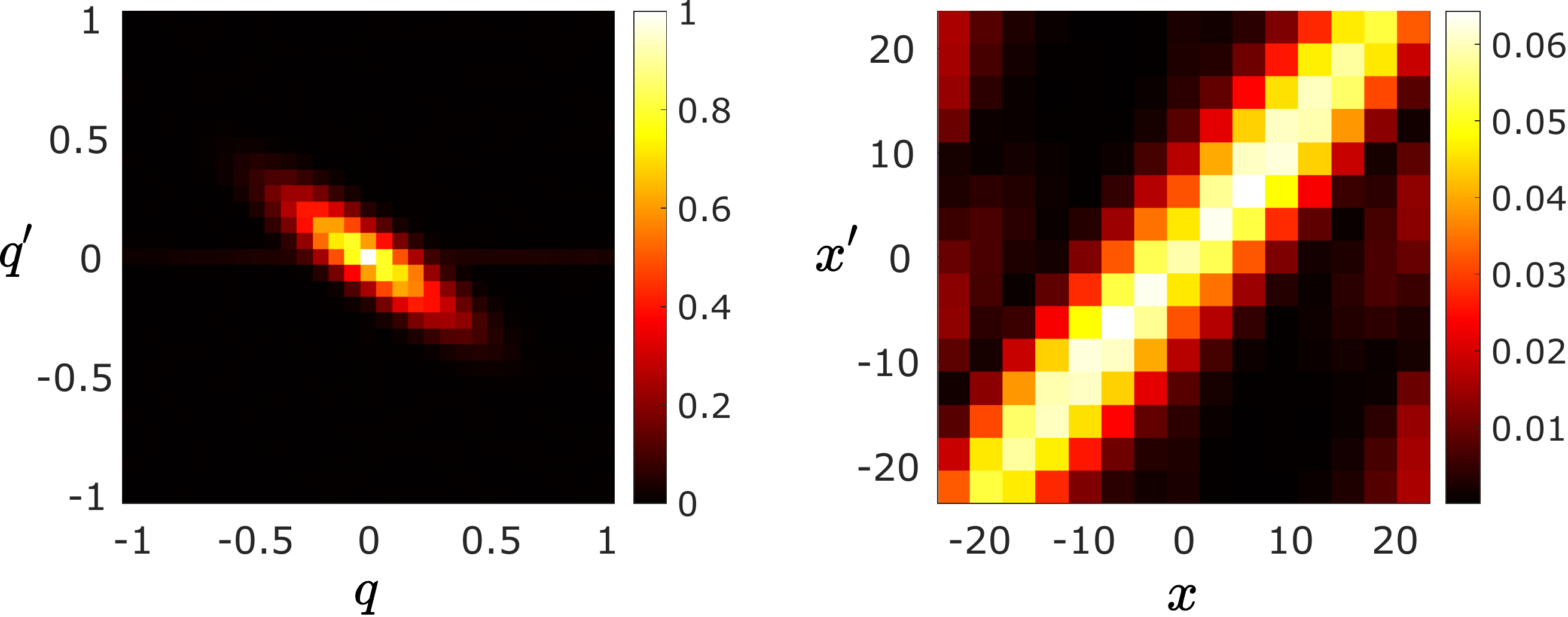}
	\end{center}
	\vspace{-14pt}\caption{Left: Signal decay $E_\Delta(q,q')$ for the spectroscopy measurement, values of $q$ and $q'$ range from $-1$ to $1$ rad/$\mu$m. Right: Center of derived propagator $P(x',\Delta|x)$, shown values of $x$ and $x'$ range from $-21.9$ to $21.9$ $\mu$m. \label{fig:spec_Results}}
\end{figure}

The complete signal decay for the spectroscopy measurement and the center of the reconstructed propagator are presented in Figure \ref{fig:spec_Results}.

\begin{figure}[h!]
	\begin{center}
		\includegraphics[width=\textwidth]{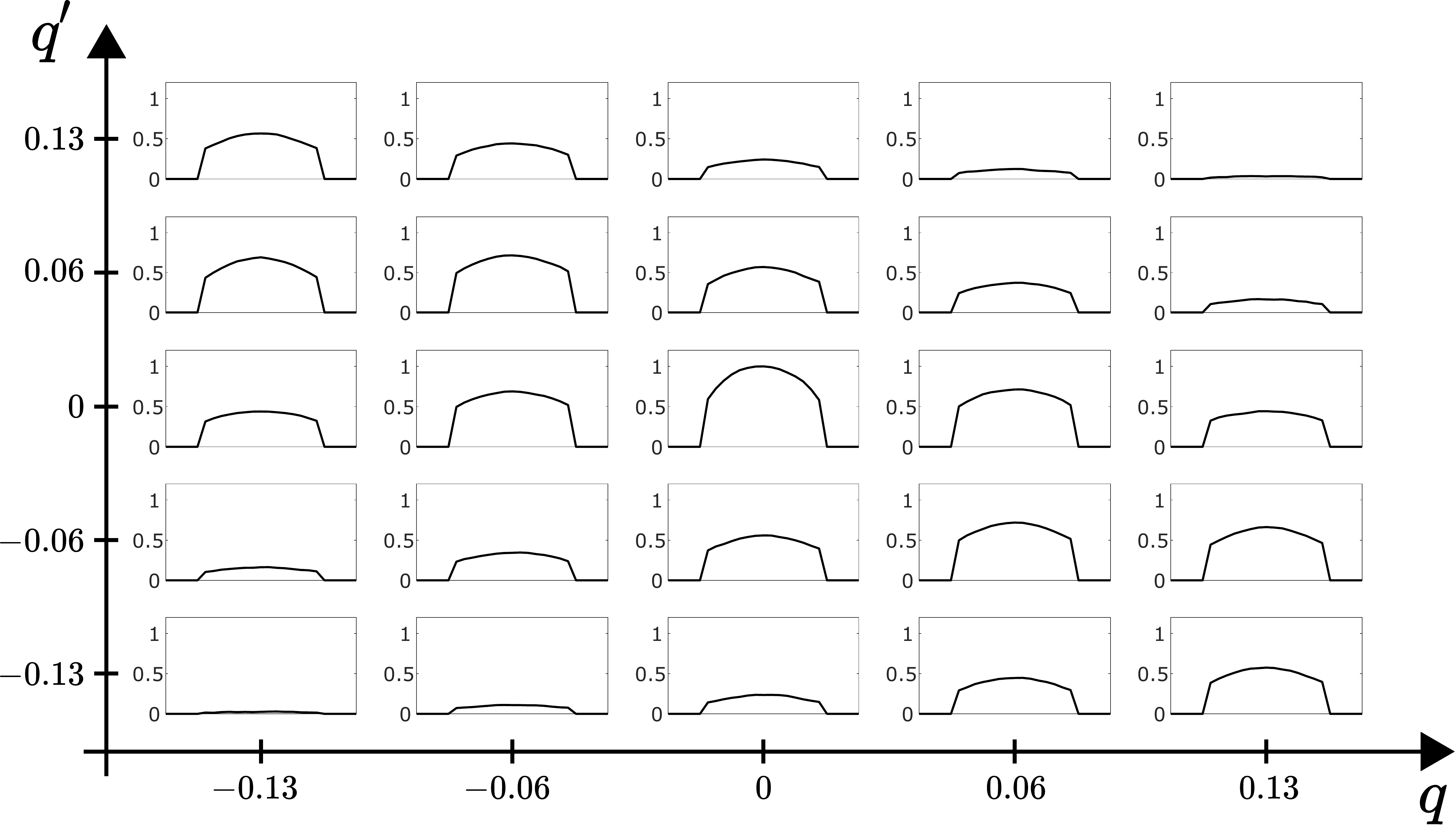}
	\end{center}
	\vspace{-14pt}\caption{Profiles obtained from the proposed sequence in a \mbox{5 $\times$ 5} matrix, each plot corresponds to a single $q$-$q'$ combination. Profiles were normalized with the magnitude of the unweighted signal's, i.e. $S_\Delta(q,q')$, middle voxel. Shown values of $q$ and $q'$ range from $-0.13$ to $0.13$ rad/$\mu$m.  \label{fig:prof_matrix}}
\end{figure}

\begin{figure}[h!]
	\begin{center}
		\includegraphics[width=\textwidth]{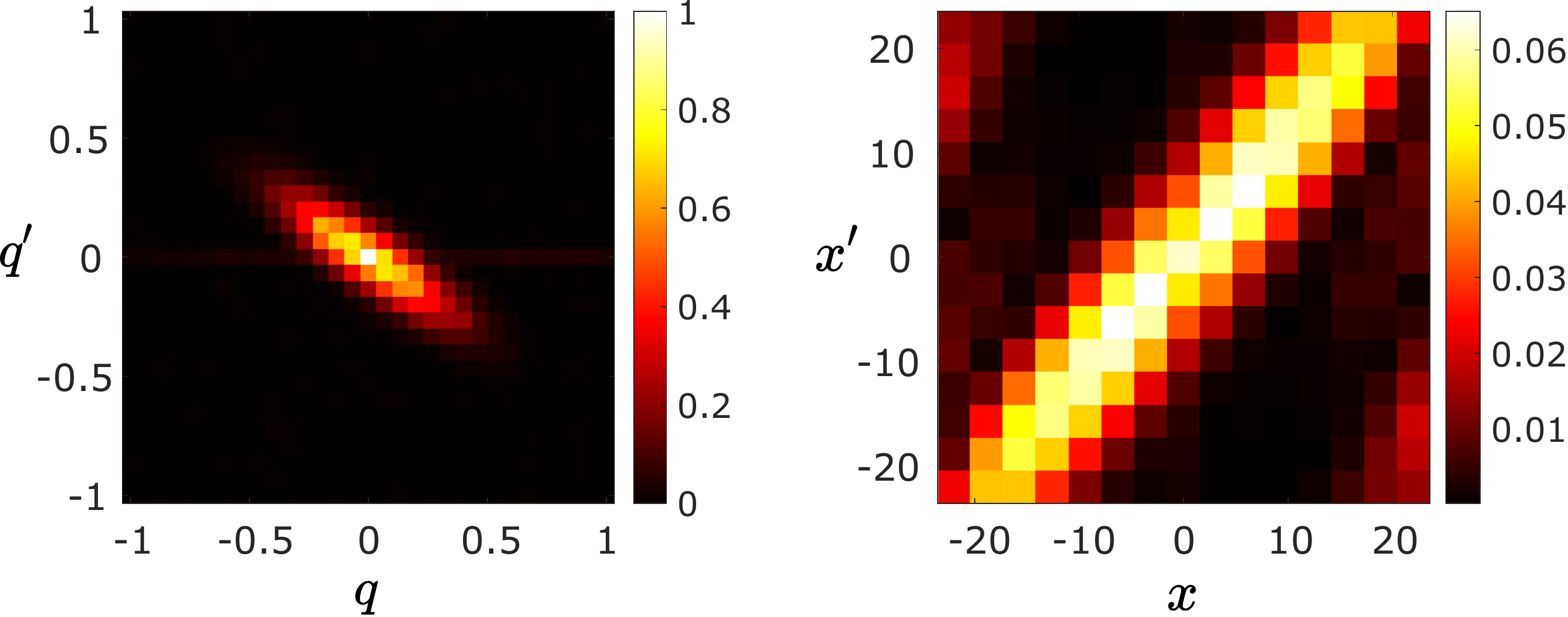}
	\end{center}
	\vspace{-14pt}\caption{Left: Signal attenuation $E_\Delta(q,q')$ for profile measurement; the values of $q$ and $q'$ range from $-1$ to $1$ rad/$\mu$m. Right: Center of the derived propagator $P(x',\Delta|x)$; shown values of $x$ and $x'$ range from $-21.9$ to $21.9$ $\mu$m. \label{fig:prof_Results}}
\end{figure}

A \mbox{5 $\times$ 5} matrix of the profiles obtained via the first experiment is presented in Figure \ref{fig:prof_matrix} where each profile was acquired with a different $q$-$q'$ combination. The complete signal decay for the middle voxel of each profile and the center of the reconstructed propagator are shown in Figure \ref{fig:prof_Results}.

\begin{figure}[h!]
	\begin{center}
		\includegraphics[width=0.97\textwidth]{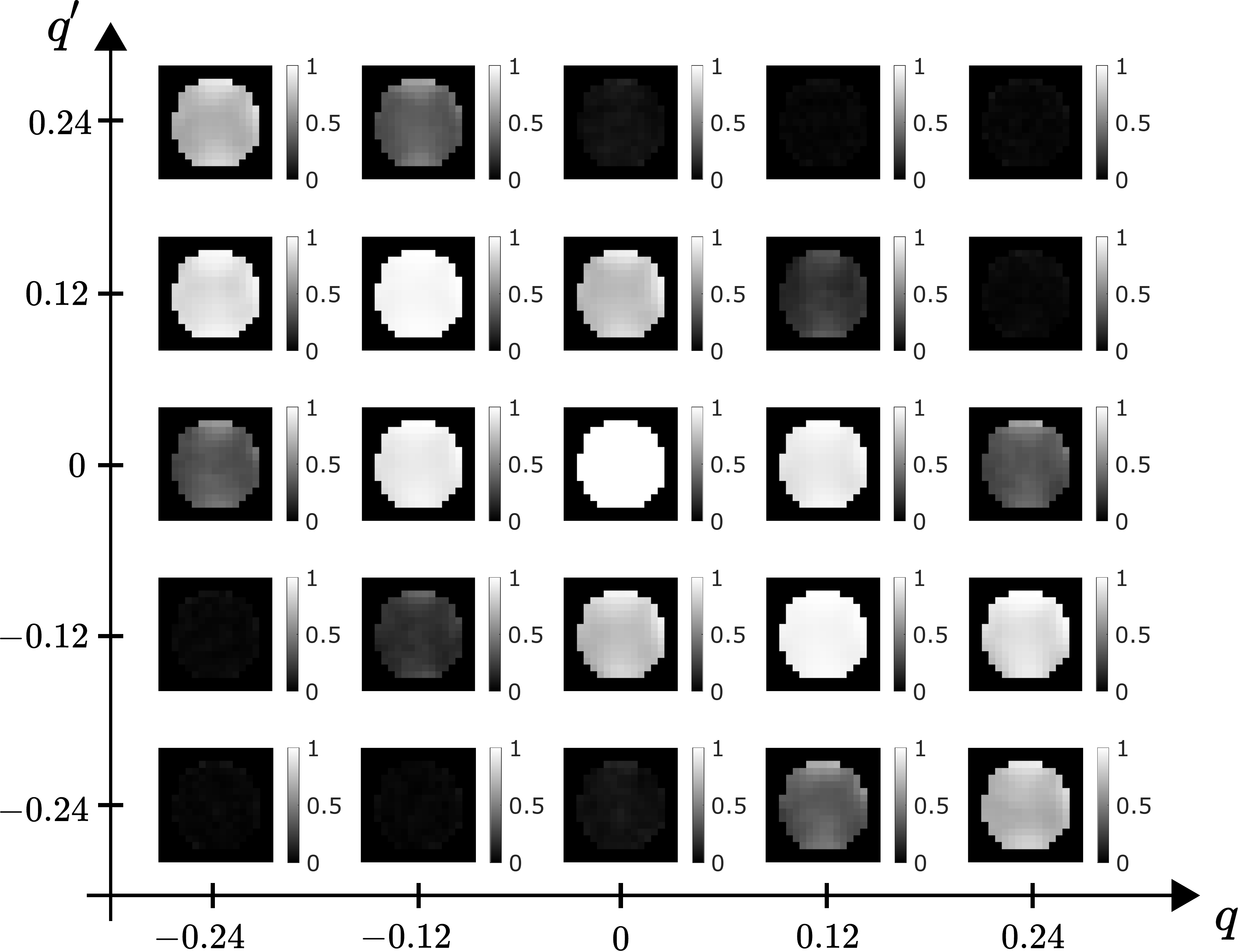}
	\end{center}
	\vspace{-14pt}\caption{Images obtained from the proposed sequence in a \mbox{5 $\times$ 5} matrix after three averages; each image is obtained with a single $q$-$q'$ combination. Images were normalized with the unweighted image, i.e., $E_\Delta(q,q')=S_\Delta(q,q')/S_\Delta(0,0)$. The plotted intensity values depict $\tanh(\pi*E_\Delta(q,q'))$ to accentuate low signal decay values at higher $q$ and $q'$. Shown values of $q$ and $q'$ range from  $-0.24$ to $0.24$ rad/$\mu$m.  \label{fig:im_matrix}}
\end{figure}

\begin{figure}[h!]
	\begin{center}
		\includegraphics[width=\textwidth]{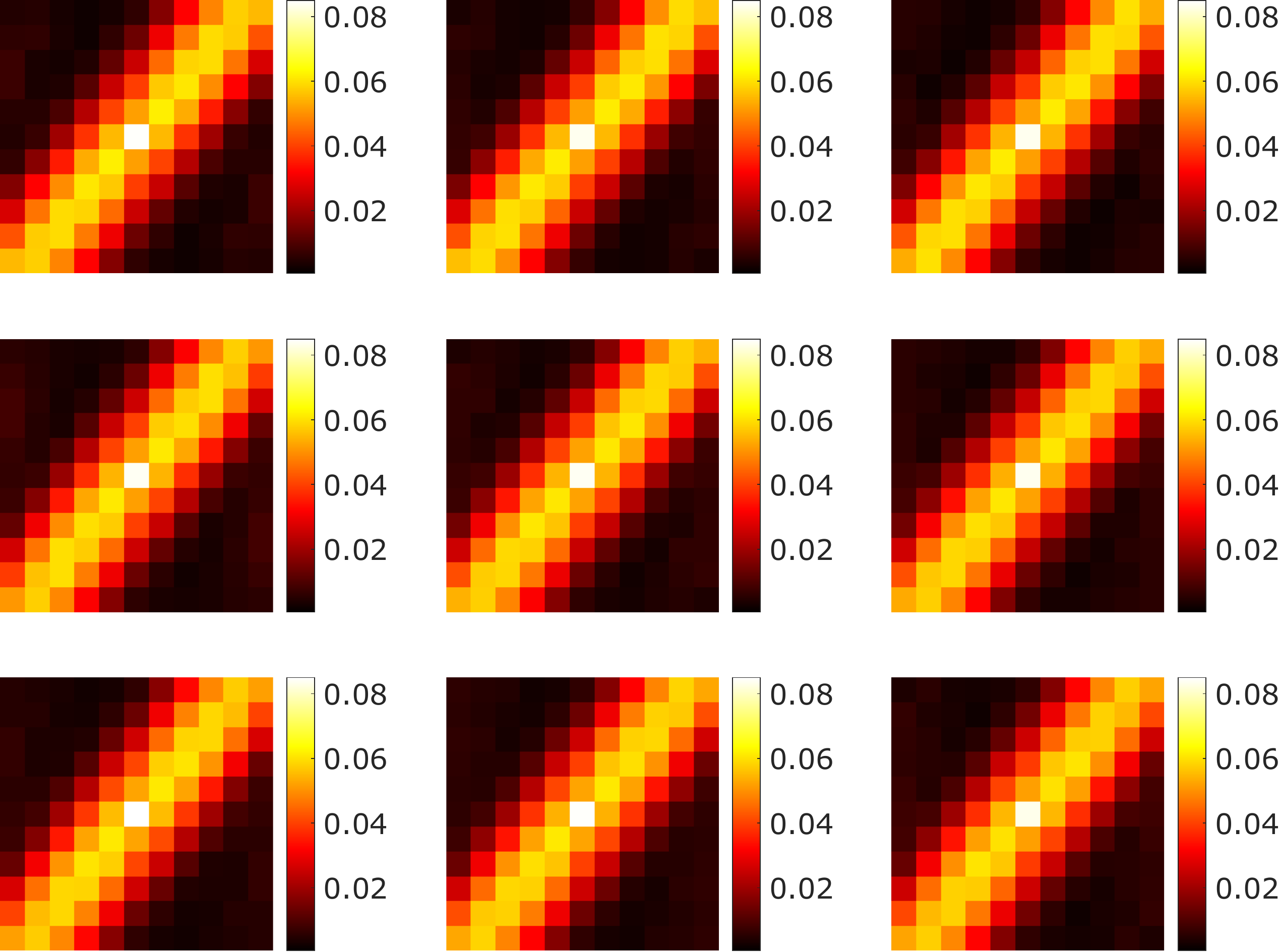}
	\end{center}
	\vspace{-14pt}\caption{Center portion of the derived propagator $P(x',\Delta|x)$ for the 9 center voxels of the acquired images in a 3$\times$3 matrix. Values of $x$ and $x'$ range from $-13.1$ to $13.1$ $\mu$m.  \label{fig:im_Pcenter}}
\end{figure}

Likewise, a \mbox{5 $\times$ 5} matrix of the normalized images acquired from the second experiment is presented in Figure \ref{fig:im_matrix}. The shown images were obtained with values of $q$ and $q'$ ranging from $-0.24$ rad/$\mu$m to $0.24$ rad/$\mu$m. The shown grayscale intensity values are related to the actual signal attenuation through the expression $\tanh(\pi*E_\Delta(q,q'))$. The center of the reconstructed propagator for the 9 center voxels are shown in Figure \ref{fig:im_Pcenter} in a \mbox{3 $\times$ 3} matrix.

\begin{figure}[h!]
	\begin{center}
		\includegraphics[width=\textwidth]{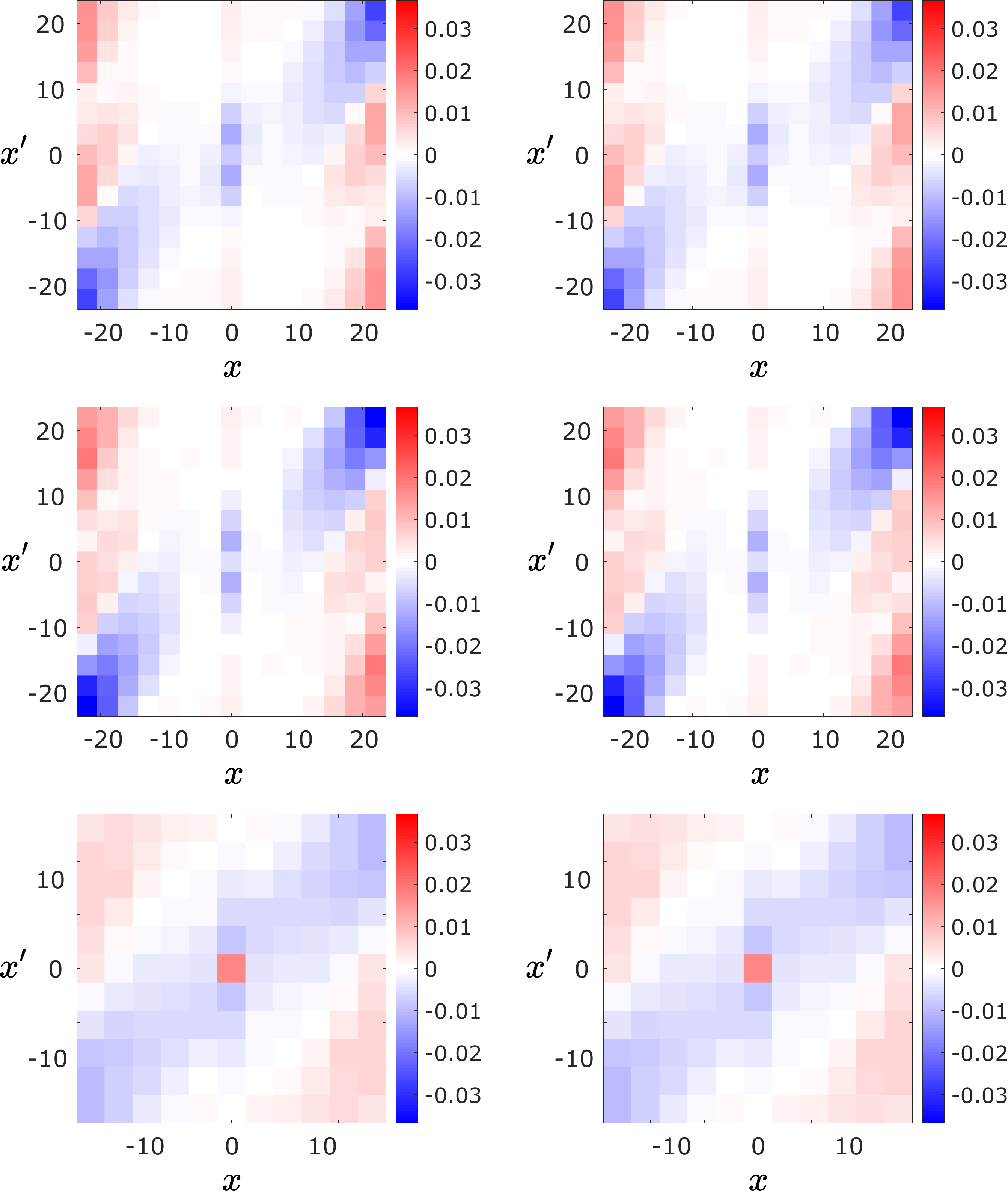}
	\end{center}
	\vspace{-14pt}\caption{Left: Difference map between the reconstructed propagator from experimental data and apparent propagator $P_\mathrm{exp}(x',\Delta | x)-P_\mathrm{app}(x',\Delta | x)$.
	Right: Difference map between the reconstructed propagator from experimental data and the one reconstructed from simulated data $P_\mathrm{exp}(x',\Delta | x)-P_\mathrm{sim}(x',\Delta | x)$. Top to bottom: spectroscopy, profiling, imaging. \label{fig:diff_all}}
\end{figure}

The difference maps between the apparent propagator obtained from equation \ref{eq:P_app} and the reconstructed one in the spectroscopy, profiling and imaging (center voxel only) experiments, as well as the difference between the propagator obtained from simulated data through equation \ref{eq:P_from_E} and the experimentally derived ones are presented in Figure \ref{fig:diff_all}. As can be observed, there is no major difference in the resulting maps corresponding to the spectroscopy and profiling measurements, especially around their center. There is, however, a noticeable difference in the center voxel of the difference maps of the propagator reconstructed with imaging data.

\begin{figure}[h!]
	\begin{center}
		\includegraphics[width=\textwidth]{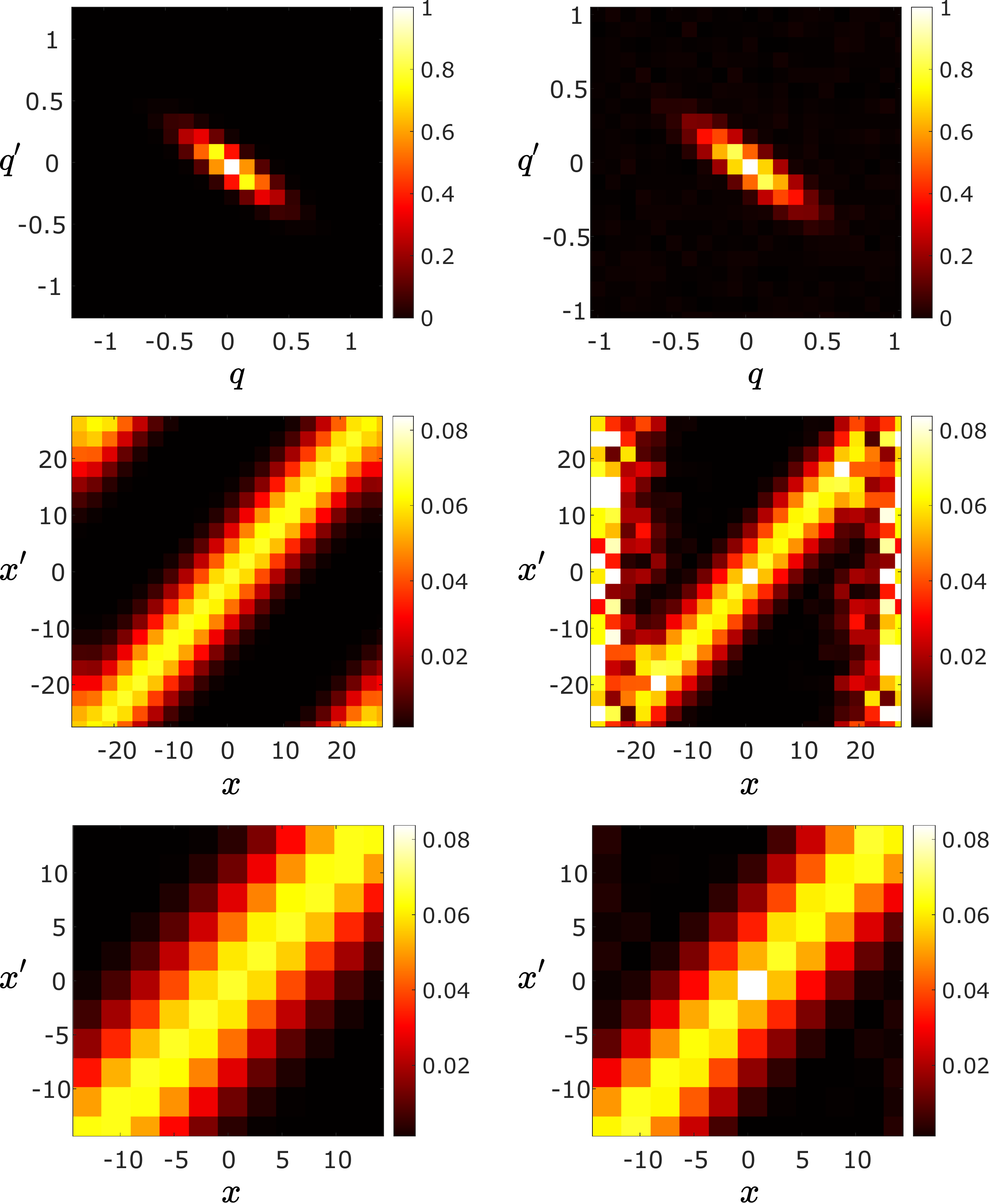}
	\end{center}
	\vspace{-14pt}\caption{Results from the simulation and propagator reconstruction for an ``ideal'' signal decay (left column) and a ``noisy'' signal decay with $\sigma = 0.01$ (right column). Top to bottom: signal decay, reconstructed propagator, center of reconstructed propagator.  \label{fig:ric_all}}
\end{figure}

The results from the simulated signal decay, the reconstructed propagator and its center are presented in Figure \ref{fig:ric_all} for both `noiseless' and `noisy'  data. Aliasing effects can be observed in the upper-left and lower-right areas of the reconstructed propagator. Furthermore, the added Rician noise results in a very noisy reconstructed propagator both at the edges of the $x$ axis and in the center voxel.

\clearpage

%%%%%%%%%%%%%%%%%%%%%%%%%%%%%%%%%%%%%%%%%%%%%%%%%%%%%%%%%%%%%%%
\section{Discussion and Conclusion}

In a recent report \cite{Ozarslan21ISMRMeverything,Ozarslan21ARXIVeverything}, a novel framework involving an acquisition sequence and an analysis technique was proposed to determine the true diffusion propagator through Fourier transforms. In this study, we proposed and implemented a modification of said sequence introducing bipolar diffusion gradients and analyzed data through simulations and experiments for the case of free diffusion.

The sequence introduced in \cite{Ozarslan21ISMRMeverything,Ozarslan21ARXIVeverything} utilizes two preferably very narrow pulses, which demand a high gradient amplitude to achieve the desired values of $q$ and  $q'$. Employing such pulses may result in the induction of eddy currents which can translate into geometric distortions in the acquired images. The bipolar diffusion gradients employed in this study reduce the effect of this type of artifacts as shown in \cite{Alexander97}. The use of the additional 180$^\circ$ radiofrequency pulses in the employed bipolar or alternating gradients also aids in the reduction of susceptibility or background gradient effects \cite{Karlicek80}.

Artifacts resulting from concomitant fields can be neglected in many acquisitions at high-fields, due to the $1/B_0$ dependence of such magnetic fields. However, the field strength of the employed benchtop scanner (0.55 T) may result in non-negligible effects which should be accounted for. When same gradient pulses having opposite polarity are applied, the quadratic terms of the magnetic field equations derived in \cite{Bernstein98} will lead to the same concomitant gradient fields. If those pulses are applied around a 180$^\circ$ radiofrequency pulse, as is the case in this study, the resulting concomitant gradient pulses within each pair will compensate each other. 

The reconstructed propagator from the three experiments performed in this study agree with the apparent propagator obtained with the employed sequence's parameters through equations \ref{eq:P_from_E} and \ref{eq:P_app}. As observed in the difference maps presented in Figure \ref{fig:diff_all}, only small variations appear in the center of the propagator, which grow slightly larger towards the corners of the $xx'$ plane.

A noticeable difference observed solely in the imaging results is the presence of a peak in the center voxel of the reconstructed propagator. This is due to employing magnitude-valued data, in which case the mean of the observed (noisy) signal is above 0 even when the underlying (noiseless) signal is 0. The presence of such `noise floor' in the signal domain is responsible for the artifactually high peak in the center of the reconstructed propagator as confirmed by the simulated `noisy' data in Figure \ref{fig:ric_all}, which exhibits the same behavior. The disappearance of this effect in the results from the spectroscopy and profiling measurements suggest that the SNR in such measurements is higher than in the imaging acquisition as expected.

Furthermore, large variations in the reconstructed propagator were observed at the edges of the $x$ axis in all experimental results as well as simulations. This was dealt with to a certain extent by discarding the edges of the propagator space with the oversampling technique. However, these variations are not desirable and should be accounted for. Such methods along with a more thorough characterization of noise introduced by the employed scanner will be considered in future studies.

With the oversampling technique employed, we managed to avoid aliasing effects in the center of the reconstructed propagator, however this increased the number of measurements needed for a proper reconstruction, which in turn yielded longer acquisition times. In this proof-of-concept study with the use of a research scanner, this is not considered to be a major issue. Nevertheless, fast imaging techniques, such as echo planar imaging (EPI), and criteria to reduce the number of required $E_\Delta(\mathbf{q},\mathbf{q}')$ samples could be of interest in future studies.

\section*{Acknowledgments}

The authors are grateful to Deneb Boito and Magnus Herberthson for their feedback on the manuscript and to Anders Eklund for his support in the acquisition of the MR scanner.

%\bibliographystyle{abbrv}
%\bibliographystyle{ieeetran}
%\bibliographystyle{apsrev4-1}
%\bibliography{\string~/Dropbox/SharedBib/sharedbib}

%\bibliography{sharedbib.bib}

%

\end{document}